\documentclass[%
 amsmath,amssymb,
preprint,%
]{revtex4-2}

\usepackage{graphicx}
\usepackage{dcolumn}
\usepackage{bm}
\usepackage{hyperref}
\usepackage{color,soul}

\begin{document}



\title{A Snapshot Review -- Fluctuations in Quantum Materials: \\From Skyrmions to Superconductivity
}

\author{L. Shen}
\affiliation{Stanford Institute for Materials and Energy Science, Stanford University, Stanford, California 94305, USA}
\affiliation{Linac Coherent Light Source, SLAC National Accelerator Laboratory, 2575 Sand Hill Road, Menlo Park, CA 94025, USA}
\affiliation{Division of Synchrotron Radiation Research, Department of Physics, Lund University, Solvegatan 14, 22100 Lund, Sweden}

\author{M. Seaberg}
\affiliation{Linac Coherent Light Source, SLAC National Accelerator Laboratory, 2575 Sand Hill Road, Menlo Park, CA 94025, USA}%

\author{E. Blackburn}
\affiliation{Division of Synchrotron Radiation Research, Department of Physics, Lund University, Solvegatan 14, 22100 Lund, Sweden}

\author{J. J. Turner}
\email{joshuat@slac.stanford.edu}
\affiliation{Stanford Institute for Materials and Energy Science, Stanford University, Stanford, California 94305, USA}
\affiliation{Linac Coherent Light Source, SLAC National Accelerator Laboratory, 2575 Sand Hill Road, Menlo Park, CA 94025, USA}%

\date{21 December, 2020}

\begin{abstract}
By measuring a linear response function directly, such as the dynamic susceptibility, one can understand fundamental material properties. However, a fresh perspective can be offered by studying fluctuations. This can be related back {to} the dynamic susceptibility through the fluctuation-dissipation theorem, which relates the fluctuations in a system to its response, an alternate route to access the physics of a material.
Here, we describe a new x-ray tool for material characterization that will offer an opportunity to uncover new physics in quantum materials using this theorem. We provide details of the method and discuss the requisite analysis techniques in order to capitalize on the potential to explore an uncharted region of phase space. This is followed by recent results on a topological chiral magnet, together with a discussion of current work in progress. We provide a perspective on future measurements planned for work in unconventional superconductivity.
\end{abstract}

\maketitle


\section{Introduction}
An important effort in material science is to continue to develop advanced characterization techniques to push the envelope of deciphering material function from microscopic properties. This is typically carried out in a myriad of ways through direct measurement of a linear response function, such as $\chi(Q,\omega)$. Another option exists for materials which is to measure spontaneous fluctuations in equilibrium, and access the linear response by making use of the fluctuation-dissipation theorem \cite{Kubo-1966-rpp}. This is a relation, first studied in relation to voltage fluctuations in circuits \cite{nyquist-1928-pr}, which relates a generalized force to properties in equilibrium. This has been utilized at high-brightness x-ray sources to access atomic resolution, most notably using x-ray photon correlation spectroscopy (XPCS) \cite{sutton-2008-crp, sinha-advmat-2014, shpyrko-2014-jsr}, but has remained challenging for timescales fast enough to connect to the relevant features in the dynamic susceptibility.

\begin{figure*}
\includegraphics[width=.5\textwidth]{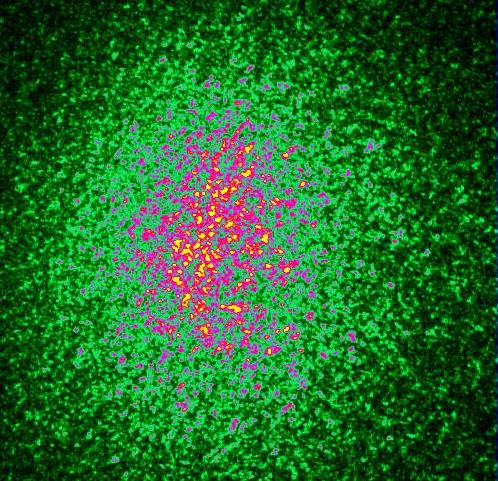}
\includegraphics[width=.475\textwidth]{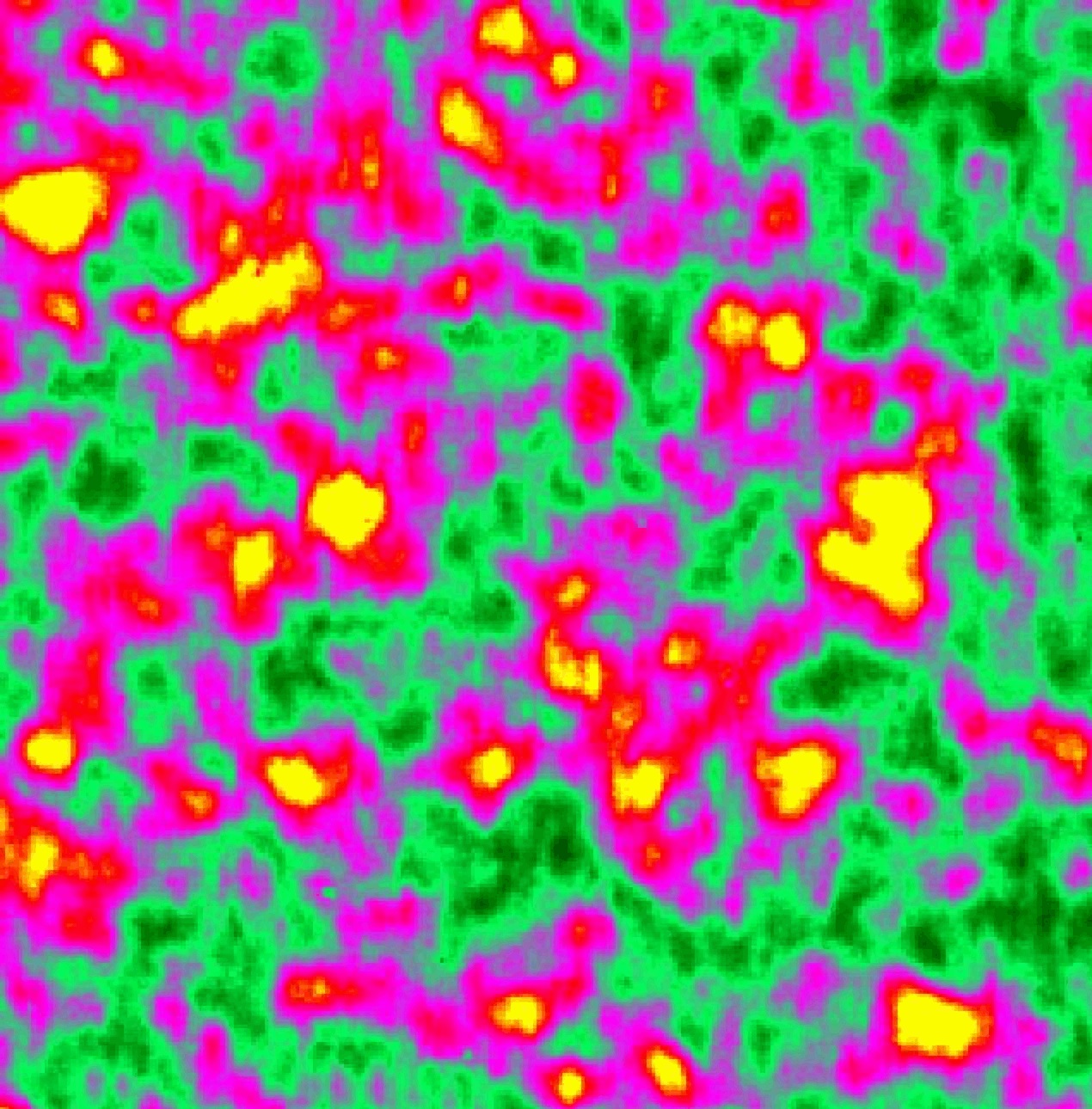}
\caption{\label{fig:1}Speckle pattern of the orbital domains in a manganite single crystal seen at the wavevector for orbital ordering. This was observed at the soft x-ray wavelengths tuned to the $L_3$-edge of Mn. The broad width of the peak is due to the short-range ordering of the orbital domains, at the level of $\sim 300\,${\AA}. Images are reproduced from \cite{Turner-2008-NJP}.
\label{fig:speck}
}
\end{figure*}

With the development of next-generation light sources such as x-ray free-electron lasers (XFELs) \cite{Emma-2010-NatPho}, new methods using coherent x-rays are becoming available at timescales of interest to the quantum materials community.
{X-ray Photon Fluctuation Spectroscopy (XPFS)} is one such method which targets the measurement of fluctuations of some underlying order parameter
in a system.  It is based on using pairs of  spatially coherent x-ray pulses, each acting as an independent data point, rather than a long pulse train as commonly used, such as in sequential XPCS. Scattering of coherent x-ray pulses from the sample generates `speckle', a complex fingerprint of structure in the sample \cite{Sutton-1991-Nature}, which is then measured and studied to monitor the dynamics. This can occur with any type of system that contains heterogeneity, such as orbital ordering domain texture for instance (see Fig.~\ref{fig:speck}).  

In its most general form, the scattering process is used in XPFS to determine the spontaneous dynamics in a system through x-ray photon fluctuations via single photon counting. XFELs provide 1.)\,fully coherent x-rays, 2.)\, femtosecond-level x-ray pulses, {and} 3.)\,a large number of photons per pulse. While these three characteristics are most well-known, there is a fourth capability required for XPFS to be achievable: the ability to create a variable two-pulse x-ray operation with a well-defined time separation. Emerging accelerator capabilities will now allow development of new tools important for the material science community. With these new capabilities \cite{decker-2015-fel, lutman-2016-natphot, lutman-2018-prl}, changes down to the level of the fs timescale will be accessible, far outpacing the state-of-the-art in XPCS at the $\mu$s timescale. This also compares favorably with inelastic neutron scattering methods, which cover the range from ps (through energy-resolved measurements) to 100s of ns and, in some cases, $\mu$s through neutron spin echo spectroscopy \cite{Gardner-2020-NRP}.

In this paper, we describe this new x-ray tool that takes advantage of the requirements above for novel materials characterization. It has the potential to extract important material parameters useful for modern condensed matter theory in solids.  We outline the XPFS method, complete with a discussion on how to perform the experiments, and delve into the analysis methods to carry out this technique on new types of quantum materials. We describe some examples with recent results on topological chiral magnets, and discuss current work in progress.  We conclude with a perspective on future measurements planned for work in unconventional superconductivity.

\section{Experimental}
We discuss the apparatus for XPFS experiments in the following section. We first give a general overview, followed by two more specific areas important for executing these types of experiments. In the first, we derive an analytic solution for the contrast function $C(q,t)$, based on the fluctuations of the scattered x-ray photons. This solution can have computational advantages for high-repetition facilities. This is followed by a general introduction to the droplet algorithm, a solution to handling large electron clouds which are created by single x-ray photons at the detector.

\subsection{XPFS: Overview}
\label{sec:overview}
When measuring dynamics, the key quantity to access, both theoretically and experimentally, is the dynamical susceptibility, $\chi(Q,\omega)$, and in particular the imaginary component of this, which is known as the dynamical structure factor $S(Q,\omega)$ {}\cite{vanhove-1954-pr}. In XPFS, information is extracted from the time domain, rather than the frequency domain, which is equally described by the intermediate scattering function $S(Q,t)$. The measured quantity in XPFS is the variance of coherent areas in the diffraction pattern, also known as the contrast function $C(Q,t)$. Importantly, this contrast function $C(Q,t)$ is related to the well-known intensity-intensity autocorrelation function $g^{(2)}(Q,t)$ \cite{Gutt-2009-OptExp} and hence the intermediate scattering function $S(Q,t)$. As these are related to the frequency domain by Fourier transformation, these type of measurements {effectively} allow direct measurement of $\chi(Q,\omega)$.


The large number of photons per pulse produced at XFELs are necessary to collect sufficient statistics from scattering at such short timescales. 
The key to the study of fluctuations in this realm is to take advantage of this, while also ensuring that the system is only weakly perturbed. The weak perturbation limit can pose a serious challenge to {perform these types of experiments}, while still supplying enough photons per pulse to conduct an XPFS experiment. {However, this} is possible due to the unique ability to incorporate large pulse energies over a short-pulse duration, the prevalent theme at XFEL sources {} \cite{pellegrini-2016-rmp}. {In this range}, the number of photons per pulse scattered from a typical sample can be only a few hundred or so, possibly a thousand, rather than many thousands or millions. This means the way to perform this type of experiment {must} change \cite{li-2014-jsr}. A new approach is required to calculate the contrast function, defined for the intensity $I(Q,t)$ as:
\begin{equation}
    C(Q,t)=\frac{I(Q,t)_{max}-I(Q,t)_{min}}{I(Q,t)_{max}+I(Q,t)_{min}} 
    \label{eq:contrast}
\end{equation}
Also known as the visibility, this describes the variance of coherent areas in the diffraction pattern. The contrast has {a} fundamental connection to the dynamic susceptibility $\chi(Q,\omega)$, which can be calculated for a given material from first principles {} \cite{green-1985-prb}. In the {single} photon limit, Eq.~\ref{eq:contrast} breaks down and instead, the XPFS method practiced in its current form involves counting x-ray photons and calculating probabilities which are related back to the requisite correlation function.

\emph{\textbf{{Coherence.}}} By starting with two fully coherent X-ray pulses, separated by some time interval $t$, we can extract a contrast from the summed speckle pattern after interaction with the sample. For negligible changes in the system during $t$, each pulse will produce an exact replica speckle pattern, while if there are significant fluctuations within $t$, the contrast of the sum will decrease. For the short timescales of interest here, both scattering patterns from each pulse of the pair need to be captured in one image. However, unraveling this is tractable since it follows an established formalism \cite{goodman-2007-book}. 

The main challenge is that in the non-perturbative limit, a small number of photons need to be used to extract an accurate contrast level $C(Q,t)$ in a single shot.  To measure $C(Q,t)$, we turn to the probability distribution of measuring $k$ photons per speckle with coherent x-rays.

\begin{figure}
\includegraphics[width=.76\textwidth]{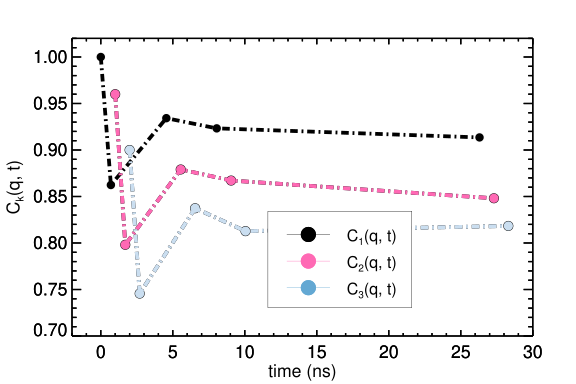}{
\caption{\label{fig:comp}Separate plots of $C_k(Q,t)$ for $k$=1-3, where each $C_k$ is based on only one probability function $P(k)$. The absolute amplitudes are slightly different for these curves but they are normalized to each other and offset, both vertically and horizontally, for comparative purposes.}
}
\end{figure}

\emph{\textbf{{Photon Fluctuations.}}} The probability distribution when using a fully coherent beam follows the Bose-Einstein distribution, the well-known probability relation for the fluctuation of bosons in a given region of phase space from statistical mechanics. Interaction with a fluctuating sample {is} analogous to imparting partial coherence to the beam, since the fluctuations will affect the detected contrast. For a partially coherent beam, the Bose-Einstein distribution is modified to approximate the negative binomial distribution. For the parameters of the experiment, the probability of detecting $k$ photons in a given speckle $P(k)$ is given by:
\begin{equation}
    P(k)=\frac{1}{k!}\prod_{\ell=0}^{{k-1}}(M + \ell)\left(\frac{{\overline{k}}}{\overline{k} + M}\right)^k\left(\frac{M}{\overline{k} + M}\right)^M 
    \label{eq:nbd}
\end{equation}
where $M$ is the degrees of freedom in the speckle pattern and $\overline{k}$ is the average number of photons per speckle \cite{goodman-2007-book}. {$M$ characterizes the amount of this effective partial coherence mentioned} above or can also be thought of as the number of coherent spatial modes contributing to the speckle, as described by Goodman \cite{goodman-1976-josa}. Using this formalism, $M$ can be extracted as a fit parameter for the measured distribution, where $M=1/C(Q,t)^{2}$.

This is equivalent to measuring $k$-photon events for a given Q, and computing the statistics for a given $t$, to extract $C(Q,t)$. We can formulate a linear combination:

\begin{equation}
    C(Q,t)=\sum_k a_k\,C_k(Q,t) 
    \label{eq:lc}
\end{equation}
to extract the information from each distribution to retrieve the final $C(Q,t)$ for $k \in \mathbb{N}$. This amounts to solving for the coefficients $a_k$. Using measurements across a range of count rates $\bar{k}$, Eq.~\ref{eq:nbd} is utilized to fit the measured distribution to obtain the resultant number of modes $M$, and finally Eq.~\ref{eq:lc} to obtain the resultant contrast. 

For example, in Fig.~\ref{fig:comp}, we plot three different $C_k(Q,t)$ values. This demonstrates that each probability $P_k(Q,t)$ has {a similar} time structure. These are normalized to each other and offset for comparison. This indicates that different portions of the data can be used separately to retrieve the same physics.

\subsection{\label{sec:analytic}Analyticity}
Given the involved {procedure} above, we have identified a simpler approach which we outline here. Because of the large number of fitting routines that need to be employed above, we can instead derive an analytic solution to extract $C(Q,t)$ from the probability distribution itself by using probability distributions ratios.

While it is clear from Eq.\,\ref{eq:nbd} that no analytic solution for $M$ exists, we can derive an analytical solution if more than one $k$-event is reliably measured. We start by taking the ratio of $P(k)$ for two different $k$-values that differ by one integer {value}. For example, by taking the ratio between $P(k+1)$ and ${P(k)}$, which we define here as ${R_k = P(k+1)/P(k)}$, the {equation can be simplified by removing terms not dependent on $k$}. By using the identity ${\Gamma(z+1)=z\Gamma(z)}$, the result is:
\begin{equation}
   R_k = \frac{k+M}{k+1} \left(1 + \frac{M}{\overline{k}}\right)^{-1}
\end{equation}
%
This simplification can be leveraged to solve for $M$ analytically. The result is:
\begin{equation}
    M = \overline{k}\frac{R_k(k+1)-k}{\overline{k}-R_k(k+1)}
    \label{eq:analytic}
\end{equation}
It should be noted that a singularity exists for $R_k=\overline{k}/(k+1)$, but this is not very probable, since $R_k$ is in general a small number, for reasonable values of $\overline{k}$. This will diverge only in cases with large numbers for $\overline{k}$ which are currently out of the range of the experimental conditions used for the systems discussed in this paper. This may need to be revisited for XPFS experiments on more robust ordered systems \cite{shen-2020-prb}, or in an area such as high-energy density sciences. For example, phase separation under dynamic compression in planetary science studies \cite{kraus-2017-natastron}, or phase boundary mapping under shock compression \cite{mcbride-2019-natphys}, are areas where there would be less constraints on $\overline{k}$ as sample damage is not a concern.

\begin{figure*}
\includegraphics[width=.6\textwidth]{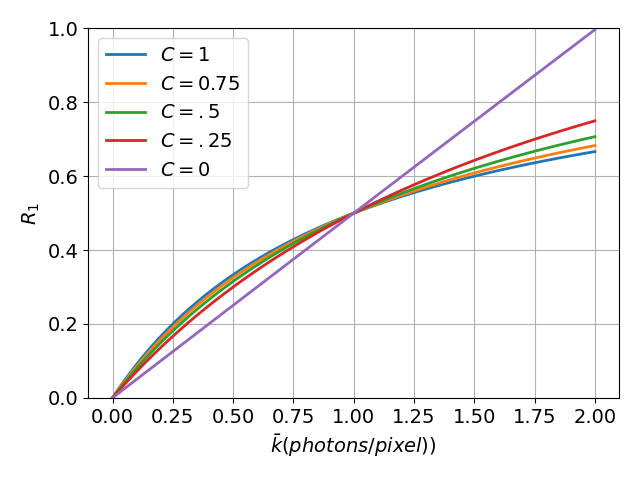}
\caption{\label{fig:c}
Contrast plot for the analytical solution derived in Eq.~\ref{eq:analytic3}. This shows $R_1$ as a function of $\overline{k}$, for given contrast values. The pinch point $R_1=1/2$ and $\overline{k}=1$ in the graph represents a singularity. This will hide the information of the contrast at this point for data collected for these parameters.
}
\end{figure*}

Equivalently, we can express this as the contrast $C(Q,t)$ which is bounded on the interval $[0,1]$ as:
\begin{equation}
    C(Q,t)=\sqrt{
    \frac{R_k(k+1)-\overline{k}}{\overline{k}[k-R_k(k+1)]}}
    \label{eq:analytic3}
\end{equation}
This last result, Eq.~\ref{eq:analytic3}, allows us to perform single-shot contrast computations. This turns the data handling into a simple problem which can be computed in real-time, a substantial advantage for high-repetition rate experiments as this development could largely speed the processing times of these types of experiments in the future.

To illustrate the behavior of the analytic solution, we  display an example of this in Fig.~\ref{fig:c}. Here{,} we plot $R_{1}$ as a function of $\overline{k}$, plotted for different values of the contrast. Note that for $k=1$, and $R_1=1/2$, a singularity exists at $\overline{k}=1$. This represents an intractable solution, where all curves intersect and no reliable value of $C(Q,t)$ can be obtained.

In conclusion, by mapping the contrast function in time and momentum space, we can extract the coefficients $a_k$ from Eq.~\ref{eq:lc} above to get $C(Q,t)$ and directly compare them to theory. The analytic solution derived here will allow for single-shot measurements which could be vital for future work at next-generation light sources.


\subsection{\label{sec:drop algo}Droplet Algorithm}

{The first step in} the data analysis in XPFS is to produce a photon map from the raw data. Because of the point-spread function which occurs with any charge-coupled detector, each photon creates an electron charge cloud. The cloud characteristics depend on the specific detector properties as well as the pixel size. For x-ray detectors with smaller pixel sizes, such as that found in the Andor system we will discuss here, the pitch can be as small as 13$\,\mu$m. {As a consequence,} a single charge cloud can cover many pixels, even for soft x-ray energies. Complications arise when clouds merge to form droplets. It is a non-trivial exercise to quantify the precise charge cloud parameters from a droplet, especially when the intensity is such that the droplet structure becomes connected. The ability of the droplet algorithm to separate the electron charge cloud within a droplet, and eventually to retrieve the photon events per speckle, is an additional constraint on limiting the pulse energy for an experiment.

\begin{figure*}
\includegraphics[width=.76\textwidth]{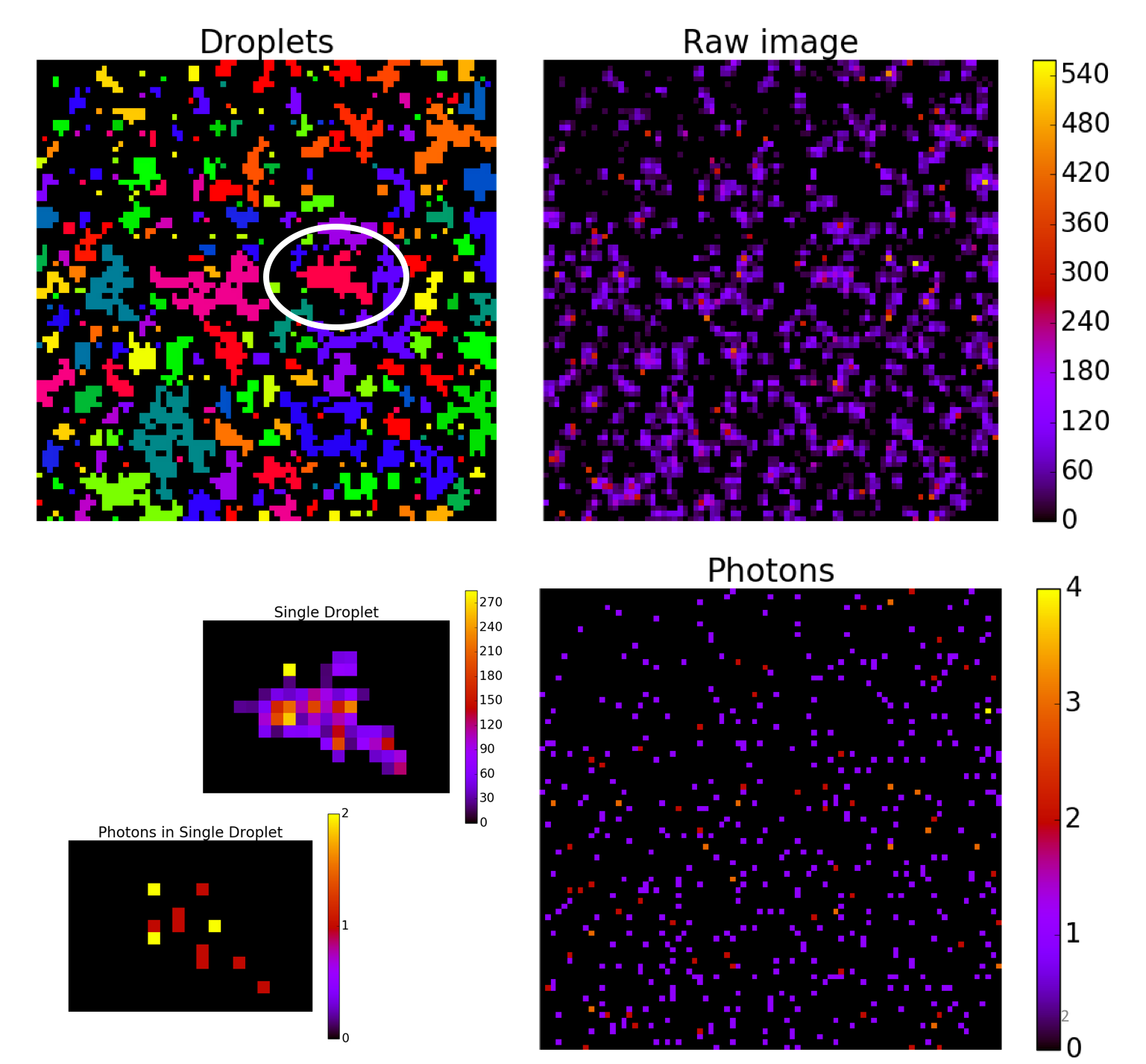}
\caption{\label{fig:2}The droplet algorithm outline, counter clockwise starting from the upper right: the raw detector image of a pair of coherent x-ray pulses scattered from a chiral magnet; The raw data showing the droplet designation for all intensity in the image, with the white circle demonstrating one single droplet; the single droplet together with it photon designation -- these are fit with multiple 2D gaussians ({Eq.}\,\ref{eq:drop}), one for each charge cloud, until the droplet is reconstructed; the full photon map of the starting image. From the gaussian fits, the measured photons per speckle can finally be resolved, recovered from the photon numbers per droplet. This is repeated for each droplet to retrieve the final photon map of the image.
}
\label{photon-map}
\end{figure*}

As an example, in Fig.~\ref{photon-map}, we outline this procedure. For each x-ray pulse pair, one coherent, summed image is collected (upper right panel in Fig.~\ref{photon-map}). By studying the topology of the resultant structure, droplets can be defined which exist as isolated structures, with no boundaries shared with other droplets (upper left panel in Fig.~\ref{photon-map}). Each droplet is assigned a number of photons based on the total analog-to-digital units (ADUs) summed across all pixels within the droplet, each of which produces its own charge cloud. The charge clouds can be modeled as a 2D gaussian ADU distribution, of the form:
%
%
\begin{equation}
    D(x, y)= \frac{N_{ph}}{\pi \sigma_x \sigma_y}{\exp}\Bigg({-{\frac{(x-x_0)^2}{\sigma_x^2}}}\Bigg) \exp\Bigg({-{\frac{(y-y_0)^2}{\sigma_y^2}}}\Bigg), 
    \label{eq:drop}
\end{equation}
where $N_{ph}$ is the expected number of ADU/photon, and $\sigma_{x}$ and $\sigma_{y}$ are the expected charge cloud widths in the horizontal and vertical direction, respectively. A first guess for the charge cloud locations can be made by recursively subtracting this distribution from the location of the global maximum within the droplet, until all photons have been accounted for. The charge cloud locations are then refined using an optimization routine, again using the gaussian model, but allowing for the charge clouds to be centered at arbitrary locations within a pixel.

As a result, each droplet can be constructed from an integer grouping of photons (lower left panel in Fig.~\ref{photon-map}). This procedure is repeated until all droplets are solved for in the diffraction image. The final result is a photon map for each speckle pattern image (lower right panel in Fig.~\ref{photon-map}). With photon maps generated for each x-ray pulse pair, we can carry out the XPFS analysis in Sec.~\ref{sec:overview} and {Sec}.~\ref{sec:analytic} to recover $C(Q,t)$.

\section{Discussion}
With the overview completed of the XPFS method above, we will now review some recent results. Though this method is still in its infancy, the progress on directly accessing magnetic fluctuations is being used to address a number of current questions in topological magnetism.

\subsection{Chiral Magnets}
Recent work has demonstrated the XPFS method discussed above to the field of chiral magnetism \cite{seaberg-2017-prl}. The first example is with the FeGd thin multilayered magnet system which has been shown to produce a skyrmion lattice. These systems are interesting because they have been shown to support skymion states in thin films at room temperature \cite{Woo2015}. The lattice properties and functionality can be tuned through the growth process \cite{Sergio-PhysRevB,Sergio2-PhysRevB}.

\begin{figure}
\includegraphics[width=.56\textwidth]{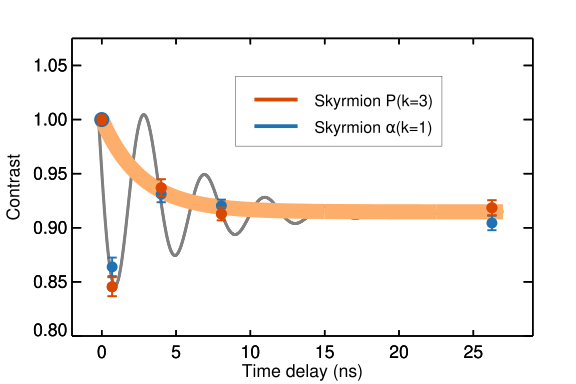}
\includegraphics[width=.38\textwidth]{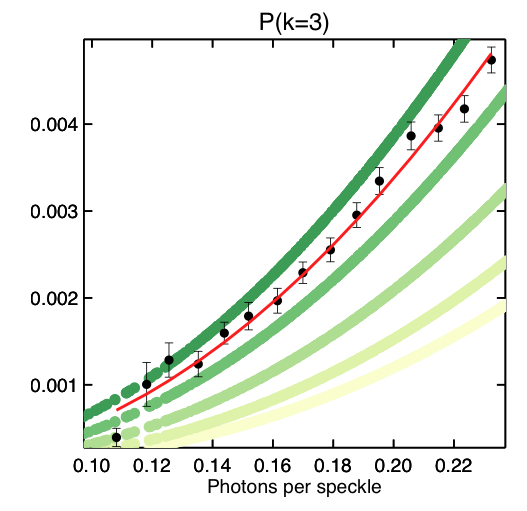}
\caption{\label{fig:5}Left: Fluctuations measured in a thin film multilayer structure of FeGd, where a skyrmion lattice forms. The long range order of the skyrmions enable detection of the lattice through soft x-ray resonant scattering with wavelength tuned to the absorption edge of the magnetic atom. Evidence of an oscillation exists, but this is being studied further. Right: an example of the probability of measuring three photons per speckle, $P(k=3)$, for the skyrmion lattice system. The fit (red line) to the data (black points) shows a contrast of 90\%, with each shade of green giving 20\% increments in contrast. This function only has one fit parameter, the number of modes $M$ discussed in the text. Reproduced from Seaberg \textit{et al.}\,\cite{seaberg-2017-prl} with permission.
}
\label{fig:old}
\end{figure}

A prototype instrument was built to execute XPFS experiments while also being able to apply a magnetic field perpendicular to the sample plane using an in-vacuum electromagnet. By stabilizing the skyrmion lattice with the field, the two-pulse data was collected on a CCD camera. The sample was measured in a forward scattering geometry and resonant x-rays were used to probe the magnetic structure. All measurements were carried out at the soft x-ray branchline (SXR) \cite{Dakovski-2015-JSR} of the Linac Coherent Light Source \cite{Bostedt-2016-RMP}, which provides monochromatic, femtosecond x-ray pulses \cite{Schlotter-2012-RSI,Heimann-2011-RSI,Tiedtke-2014-OptExpress,Chalupsky-2011-NIMPR,Krupin-2012-OptExp,Beye-2012-APL,zohar-2019-ol}.

\begin{figure}
\includegraphics[width=.96\textwidth]{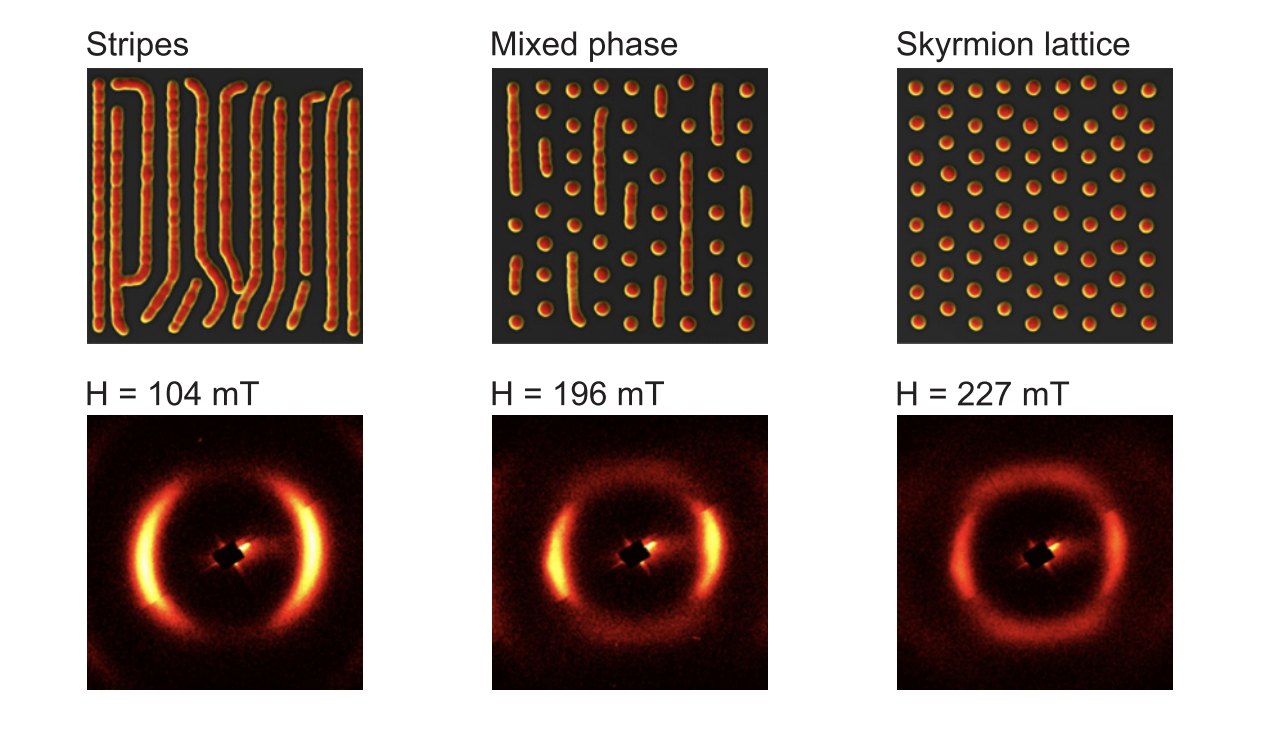}
\caption{\label{fig:6}Fe–Gd thin film heterostructure phases at room temperature. Top: Sketch of the stripe, mixed, and skyrmion
lattice phases in real space, based on simulations from previous work \cite{Sergio-PhysRevB,Sergio2-PhysRevB}. Bottom: Measured soft XFEL data of the  diffraction patterns at the Gd $M_5$-edge showing
the concomitant phases in reciprocal space. Data reproduced from Esposito \textit{et al.}\,\cite{esposito-2020-apl}, with the permission of AIP Publishing.
}
\label{apl}
\end{figure}

The data revealed pure exponential relaxation on the order of a couple of nanoseconds shown in Fig.~\ref{fig:old}, potentially indicating an oscillation, but this needs further investigation. These represent the spontaneous fluctuations inherent to the skyrmion lattice, and yield new insight into $\chi(Q,\omega)$ for these types of topological systems. The strength of this measurement is not only the observation of the fluctuation spectrum, but the information it yields on the skyrmion-skyrmion interaction, as well as the dynamic fraction of the skyrmion lattice. The contrast curve in Fig.~\ref{fig:old} indicates that only about one-third of the lattice structure {is} fluctuating on this timescale. The theoretical work in this area is ongoing.

Furthermore, future work that involves randomly coded masks and numerical convex relaxation routines \cite{seaberg-2015-apl,waldspurger-2015-mathprog,fogel-2016-mpc} may be able to produce single-shot images of systems such as this, in this geometry \cite{Miao-2015-science}. This direction would necessitate more photons and would be in the perturbative regime, usually at the cost of sample damage. This area is still under investigation for future high-repetition rate XFEL machines.

\subsection{The Biskyrmion Lattice}
Another area which has been explored by XPFS is in the study of biskrymion lattice structures, {or skyrmion bound pairs}. The sample was again a FeGd chiral magnet, but with certain field geometry and ramping procedure \cite{Lee-2016-APL}, a biskyrmion lattice state can be created. These objects consist of two bound skyrmion objects, which {can act} together as a separate quasiparticle. They have been the focus of recent work which has shown that they can be controlled with small magnetic fields, and with many orders of magnitude of smaller fields, for advanced technologies \cite{yu-2011-natmat}.


For this study, we constructed a more advanced instrument, installed a pixelated large area detector, and added a contrast monitor, which can monitor the single pulse contrast for normalization. This instrument also had the UHV electromagnet for perpendicular magnetic field stabilization \emph{in-situ} during the x-ray measurement and was installed at the SXR instrument \cite{Dakovski-2015-JSR}.

In this example, we explored the fluctuations near the first-order phase transition. In Fig.~\ref{apl}, images of the real space figuration are shown for the ferromagnetic stripe phase, the skrymion lattice phase, and the mixed phase at the boundary between the two, with each point in the diagram representing one biskyrmion object. Below these images in the figure are the respective diffraction patterns which indicate which phase is under study. These are all room temperature phases that are accessed through magnetic field application. The scattering used for detection is resonant soft x-ray scattering from the Gd atoms at the $M_5$-edge.

\emph{\textbf{Skyrmion Glass.}}
We used XPFS to search {for} fluctuations at the first order boundary. In contrast to the behavior in the example found above, we observed the dynamics to exhibit a Kohlrausch–Williams–Watt-type decay:

\begin{equation}
    C(Q,t)=C_0 + C_1\exp^{-{[ t(Q)/\tau]}^\beta} 
    \label{droplet}
\end{equation}
with significant stretching, given by the parameter $\beta$ \cite{esposito-2020-apl}. This direct measurement of the magnetic fluctuations, near the
coexistence region of the discontinuous phase transition, was interpreted as
indicative of jamming or `glassiness'.
For the former, the theory is that the limited available volume for the biskrymion lattice phase is set by the ferromagnetic stripe phase in the phase coexistence region which causes limited, i.e. non-gaussian dynamics, which
would indicate jamming. Stretched exponential dynamical behavior has also been seen in different types of glasses \cite{phillips-1996-rpp}, but further measurements are needed to provide definitive proof on {whether} this phase represents a true `skyrmion glass' or not.

\section{Perspective}
In this section, we discuss ongoing work and future developments. We will first discuss the scientific topics currently being addressed with XPFS followed by an outlook on important developments that will be key to the success of XPFS: advanced instruments and detectors.

\subsection{Fluctuating Order in Unconventional Superconductors}

One of the interesting topics in high-temperature superconductivity is in the observation of density wave order, both charge (CDW) and spin (SDW), and its relationship to the mechanism of superconductivity. This was first discovered using elastic neutron scattering 
in superconducting La$_{2-x}$Sr$_{x}$CuO$_{4}$ \cite{cheong-1991-prl} and after a series of recent x-ray measurements \cite{Ghiringhelli-2012-science,achkar-2012-prl, chang-2012-natphys}, is now understood to be a ubiquitous feature of these systems. This has led to a revitalized focus on these materials for understanding how CDWs and SDWs, sometimes generally referred to as `stripes', relate to the mechanism of high-temperature superconductivity \cite{emery-1999-pnas}. Closely related to the CDW state is the pseudogap phase \cite{dai-2001-prb}, which refers to an energy scale at the Fermi level which has a decrease in density of states associated with it \cite{Hufner-2008-rpp}. 

A myriad of papers have now been devoted to revealing the stripe fluctuations in cuprate superconductors, using inelastic neutron scattering (INS) \cite{Tranquada2004,Vignolle2007}, resonant inelastic x-ray scattering (RIXS) \cite{LeTacon-2011-NP,miao-2017-pnas,Hepting2018,Arpaia906}, and pump-probe spectroscopy \cite{torch-2013-natphys, hinton-2013-prb, dakovski-2015-prb,mitrano-2019-sciadv, wadel-2020-arxiv}. Among these techniques, INS and RIXS are {most} powerful in measuring the coherent excitations in solids, while pump-probe methods normally evaluate the non-equilibrium states transiently created by a femtosecond optical pulse.

One piece of missing information, however, is the measurement of fluctuations in equilibrium over the full range of timescales.  It is the essence of these quantum stripes that could be relevant for understanding the superconducting mechanism \cite{kivelson-2003-rmp,RevModPhys.87.457}, and which may even $\mathit{enhance}$ the superconducting state. Theoretically, it has been shown that the stripe instability vanishes and superconductivity becomes stronger when transverse stripe fluctuations have a sizable amplitude \cite{kivelson-1998-nature}.

The stripe fluctuations have been extensively searched for in experiment. Taking the canonical cuprate La$_{2-x}$Ba$_{x}$CuO$_{4}$ ($\it{x}$~=~1/8) for instance, a recent femtosecond time-resolved resonant soft x-ray scattering (pump-probe) investigation at the LCLS reported the absence of a photo-induced coherent mode \cite{mitrano-2019-sciadv}. This observation rules out the static charge stripe scenario in this system and indicates that the fluctuations must be gapless, i.e.\,relaxational in the time domain. RIXS has also been utilized to explore stripe fluctuations therein, which revealed that the fluctuations are slower than 100~fs \cite{miao-2017-pnas}. The time window of interest has been further narrowed down by XPCS, in which static charge stripes at the second to hour timescale have been established \cite{chen-2016-prl}. Consequently, the ps to ms regime is the `sweet spot' to finally unravel the fluctuations in this compound. Interestingly, this `sweet spot' appears to coincide with a very recent study on this compound, which reports possible fluctuations at the NMR timescale: $\sim$ 1$\mu$eV or 5~nanoseconds \cite{PhysRevB.101.174508}. All these results taken together imply that nanosecond XPFS will be a unique tool to explore the relevant physics in this area.


 The description of the XPFS method outlined earlier provides a solution uniquely matched to this problem.   
Taking advantage of the spatial coherence of the x-ray beam available at XFELs, speckle measurements will be sensitive to the stripe fluctuations in equilibrium. Current work is ongoing and aimed at shedding new light on high-temperature superconductivity by studying these dynamic stripe fluctuations in the time domain.

Additionally, because the recently discovered CDW peak in the bulk is short-range ordered, the state could produce `liquid crystal-like' fluctuations, which has also been described with recent theoretical work \cite{markiewicz-2016-scirep}. Here{,} it was shown that a smectic-nematic phase could be described by a CDW coupled to strain, analogous to measurements which have used coherent x-rays to capture over-damped liquid crystal fluctuations \cite{price-1999-prl}. Because electronic liquid crystal phases can also be shown to be analogous to the stripe phase \cite{emery-1999-pnas}, focusing on measurements  to unravel potential new physics via pure CDW dynamics is also promising. Furthermore, another approach is to treat the electronic correlations with an intermediate coupling model. A framework for obtaining properties using a correction for the self-energy which incorporates correlation effects is also a reassuring formalism that might address these types of measurements \cite{markiewicz-scirep-2017}.

\begin{figure*}
\includegraphics[width=.96\textwidth]{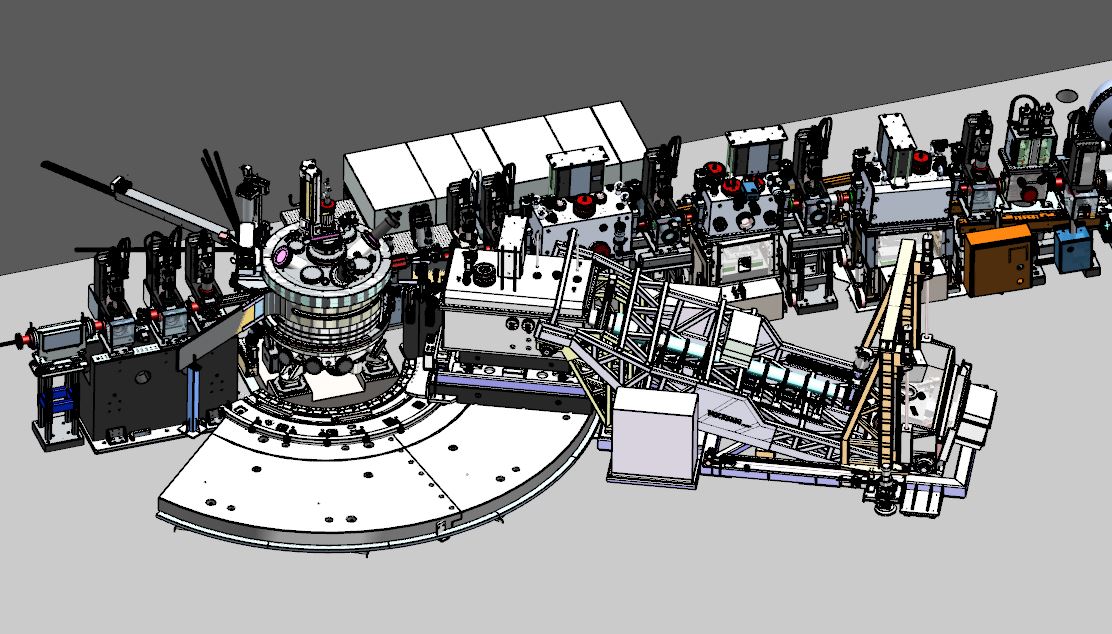}
\caption{The so-called qRIXS instrument design, planned for installation at the LCLS-II facility. This instrument is being built for time- and momentum-resolved resonant inelastic x-ray scattering measurements in the soft x-ray regime, but will be ideally suited for XPFS. The circular sample chamber holds the sample and both the spectrometer and detector arm is capable of moving through a scattering angle from 40$^\circ$ to 150$^\circ$. A detector will be placed at a given distance from the sample chamber and can be placed at an arbitrary scattering angle within this range using the RIXS mechanical infrastructure. This detector will directly capture the speckle for the XPFS measurement. (Image provided by Jim Defever and Frank O'Dowd; SLAC National Accelerator Laboratory).}
\label{qrixs}
\end{figure*}

\subsection{Advanced Instruments}
Currently under development at the LCLS-II facility is the so-called qRIXS endstation. This is under construction and will take advantage of the high repetition rate of the LCLS-II machine in the soft x-ray range (See Fig.~\ref{qrixs}). The design of this advanced instrument will make it the ideal location to perform future XPFS measurements.

This machine is being built for time- and momentum-resolved resonant inelastic x-ray scattering measurements and includes a sample environment chamber to cool and orient the sample together with a spectrometer and detector arm which has a motion covering scattering angles from 40$^\circ$ to 150$^\circ$. This will detect the scattered x-ray spectrum on a movable detector, including polarization analysis.

\emph{\textbf{Future XPFS.}}
To conduct XPFS measurements, an advanced detector (See Sec.~\ref{axd}) will be located at a given distance from the sample chamber and can be positioned at an arbitrary scattering angle within this range using the RIXS mechanical infrastructure. This detector will directly capture the speckle-diffraction for the XPFS measurement. The detector will be placed in the scattering plane and needs to have the characteristics to resolve {the} speckle {structure}, while also being able to be read out at high repetition rates.

\subsection{Advanced X-ray Detectors}
\label{axd}
Given the discussion above, there are a few critical needs for XPFS that we have discussed: short bursts of high pulse energy x-rays, with a capability to adjust the timing between pulses on the requisite timescale, and spatial coherence. Technically, another key element is development of next-generation detectors for XPFS. Since this infrastructure is critical to the needs of the materials community, we outline some points on this development here. Noteworthy issues, in no particular order, are the following: (1) Repetition rate, (2) pixel size, (3) a photon-counting mode, (4) charge-sharing characteristics, and (5) array size. We also describe the latest development on promising candidates for the next XPFS detector.

\emph{\textbf{Repetition Rate.}} 
With LCLS-II capabilities reaching MHz repetition rates, the read-out rate of the detector is critical to take advantage of this progress. Because XPFS is based on {statistics of a small number of photons in a given shot}, it is perfectly suited for high-repetition rate data collection. XPFS will immensely benefit from high-repetition rates of the machine, and so any deficiency in detector repetition rate capability will directly translate to a loss of x-ray data. 

We can estimate the {pulse pair image} conditions at 1\,MHz. Starting with a maximum of 120\,kW of electron beam power from the accelerator, with 4\,GeV at 1\,MHz gives 30\,pC, or about 15\,pC for each pulse of the pulse pair. Under typical LCLS conditions, a $\sim$ 2\,mJ/pulse is generated for a 250\,pC bunch pulse, or for 15\,pC per pulse, we expect more like $\sim$ 0.1\,mJ/pulse. Assuming similar performance of the optics used at the {SXR} branchline at LCLS {}\cite{Soufli-2012-AppOpt,Moeller-2015-JSR}, we expect about $\sim 0.2\%$ efficiency, depending on bandwidth \cite{Tiedtke-2014-OptExpress}, or about 200\,nJ/pulse on the sample. In the previous conditions for the chiral magnetic films, we applied about $\sim$ 1\,nJ/pulse to the sample, after heavy attenuation for a 30\,$\mu$m spot. For this particular case, we saw unwanted changes at around 50\,nJ/pulse, which means under a similar experimental configuration, experiments would need to use attenuation on the order of about a factor of $\sim$ 200.

\emph{\textbf{{Pixel Size.}}}
XPFS is based on formulating a speckle pattern (see Fig. \ref{fig:speck}). This is formed from interference due to scattering from a strongly heterogeneous sample as a result of a spatially coherent x-ray beam illumination together with a heterogeneous length scale which is smaller than the coherence length of the beam and larger than the wavelength of x-rays being used \cite{dainty-1985-book}. This technique works best by detecting a given speckle, which sets the limit on the pixel size of the detector: ideally, the pixel must be smaller than or equal to one speckle, where there may be added benefit to having multiple pixels per speckle. Since XPFS is based on measuring statistics within a speckle, resolving a speckle is important and with it, having a small pixel size. This size directly sets the geometry of the experiment, and hence, the size of the instrument to measure it.

\emph{\textbf{{Single Photon Counting Mode.}}}
XPFS is based on photon counting, and hence needs a detector that counts single photons {reliably} above the noise level. Additionally, by definition, 2-photon and 3-photon events are needed as well. {This means that sensitive, high-gain detectors which can differentiate a 2-photon and a 3-photon event per speckle are vital}.

\emph{\textbf{Charge Sharing Characteristics.}}   
Because XPFS is dependent on counting photons, each photon event must be able to be reliably recognized. This means that charge sharing between pixels needs to be corrected for. Currently, to account for this, droplet algorithms \cite{Hruszkewycz-2012-PRL} need to be implemented, which provide a way of fitting the distribution of intensity measurements on the detector and extrapolating the photon events that occur for each pixel (See Sec.~\ref{sec:drop algo}). This can be time-consuming, and challenging, and can greatly hinder these types of measurements going into the future, so new XPFS detectors will need to minimize this. 

\emph{\textbf{{Array Size.}}}
{Maximizing} the full array size is always beneficial, but this is also particularly tied to the specific experiment. {This is important both} for the scattered signal characteristics and the geometry, the latter setting the speckle size, and hence the distance of the detector from the sample. {In other words, the pixels should be designed to be small enough to resolve the speckle structure, while allowing the full array size to be optimized to collect the maximum speckle number for a given diffraction distribution}.

\emph{\textbf{Detector Development.}}
The long term development plan at the SLAC National Accelerator Laboratory is to build experiment specific x-ray cameras, the so-called `SparkPix' family. This is centered around the idea that for some class of experiments, one can use the structure or the characteristics of the data to extract efficiently the information in real time inside the detector itself. {This} family is built as an extension of the ePix family \cite{Sikorski2016}, reusing all ePix core circuit modules combined with different kinds of information extraction engines. These are set to extract information at the full repetition rate of the LCLS-II at 1\,MHz, or in burst mode at 3\,GHz and includes advanced extraction features such as sophisticated triggering and zero suppression. 

Most critical for XPFS is the idea of sparsification. Given the discussion above, it is clear that a typical XPFS experiment will only have a small number of pixels illuminated on a given shot. To produce a full-frame readout is unnecessary. The sparsification engine would be a layer in these type of advanced detectors which would only read out pixels which detect an event. This will allow extremely high repetition rates, while collecting the full available data. These types of developments are what will make the XPFS method valuable in the materials science community looking forward.

\section{Conclusion}

With the latest state-of-the-art developments at XFELs, new characterization tools are becoming available which could have a dramatic impact on understanding the properties of matter. XPFS is a new method which allows access to a region of phase space that so far is uncharted territory in the x-ray regime. While work has started in chiral magnetic systems, the future is bright with technical developments{,} such as in next-generation x-ray facilities and instruments{, as well as} in new technological advances in fast, large-area detectors. With these in mind, new classes of experiments to comprehend fluctuating order and how this is related to unconventional superconductivity is at our fingertips.

\section{Data availability statement}
The datasets generated during and/or analysed during the current study are available from the corresponding author on reasonable request.

\begin{acknowledgments}
We gratefully acknowledge discussions with A. 
Dragone, F. O'Dowd, J. Defever, and G. Dakovski. This work is supported by the U. S. Department of Energy, Office of Science, Basic Energy Sciences, Materials Sciences and Engineering Division, under Contract DE-AC02-76SF00515. Use of the Linac Coherent Light Source, SLAC National Accelerator Laboratory, is supported by the U.S. Department of Energy, Office of Science, Office of Basic Energy Sciences under Contract No. DE-AC02-76SF00515.
J. J. Turner acknowledges support from the U.S. DOE, Office of Science, Basic Energy Sciences through the Early Career Research Program.

\end{acknowledgments}



\providecommand{\noopsort}[1]{}\providecommand{\singleletter}[1]{#1}


\begin{thebibliography}{10}

\bibitem{Kubo-1966-rpp}
R~Kubo.
\newblock The fluctuation-dissipation theorem.
\newblock {\em Reports on Progress in Physics}, 29(1):255--284, jan 1966.

\bibitem{nyquist-1928-pr}
H.~Nyquist.
\newblock Thermal agitation of electric charge in conductors.
\newblock {\em Phys. Rev.}, 32:110--113, Jul 1928.

\bibitem{sutton-2008-crp}
M.~Sutton.
\newblock A review of x-ray intensity fluctuation spectroscopy.
\newblock {\em C. R. Physique}, 9:657, 2008.

\bibitem{sinha-advmat-2014}
Sunil~K. Sinha, Zhang Jiang, and Laurence~B. Lurio.
\newblock X-ray photon correlation spectroscopy studies of surfaces and thin
  films.
\newblock {\em Advanced Materials}, 26(46):7764--7785, 2014.

\bibitem{shpyrko-2014-jsr}
Oleg~G. Shpyrko.
\newblock {X-ray photon correlation spectroscopy}.
\newblock {\em Journal of Synchrotron Radiation}, 21(5):1057--1064, Sep 2014.

\bibitem{Emma-2010-NatPho}
P.~Emma, R.~Akre, J.~Arthur, R.~Bionta, C.~Bostedt, J.~Bozek, A.~Brachmann,
  P.~Bucksbaum, R.~Coffee, F.~J. Decker, et~al.
\newblock First lasing and operation of an ångstrom-wavelength free-electron
  laser.
\newblock {\em Nat. Photonics}, 4(9):641--647, 2010.

\bibitem{Sutton-1991-Nature}
M.~Sutton, S.~G.~J. Mochrie, T.~Greytak, S.~E. Nagler, L.~E. Berman, G.~A.
  Held, and G.~B. Stephenson.
\newblock Observation of speckle by diffraction with coherent x-rays.
\newblock {\em Nature}, 352(6336):608--610, 08 1991.

\bibitem{decker-2015-fel}
F~Decker, S~Gilevich, Z~Huang, H~Loos, A~Marinelli, C~A Stan, J~L Turner, Z~Van
  Hoover, and S~Vetter.
\newblock {Two Bunches with ns-separation with LCLS}.
\newblock In {\em Proceedings of FEL2015}, pages 634--638, 2015.

\bibitem{lutman-2016-natphot}
Alberto~A. Lutman, Timothy~J. Maxwell, James~P. MacArthur, Marc~W. Guetg, Nora
  Berrah, Ryan~N. Coffee, Yuantao Ding, Zhirong Huang, Agostino Marinelli,
  Stefan Moeller, and Johann C.~U. Zemella.
\newblock Fresh-slice multi-colour x-ray free-electron lasers.
\newblock {\em Nature Photonics}, 10:745 EP --, 10 2016.

\bibitem{lutman-2018-prl}
Alberto~A. Lutman, Marc~W. Guetg, Timothy~J. Maxwell, James~P. MacArthur,
  Yuantao Ding, Claudio Emma, Jacek Krzywinski, Agostino Marinelli, and Zhirong
  Huang.
\newblock High-power femtosecond soft x rays from fresh-slice multistage
  free-electron lasers.
\newblock {\em Phys. Rev. Lett.}, 120:264801, Jun 2018.

\bibitem{Gardner-2020-NRP}
Jason~S. Gardner, Georg Ehlers, Antonio Faraone, and Victoria Garcia~Sakai.
\newblock High-resolution neutron spectroscopy using backscattering and neutron
  spin-echo spectrometers in soft and hard condensed matter.
\newblock {\em Nature Reviews Physics}, 2:103--116, January 2020.

\bibitem{Turner-2008-NJP}
J~J Turner, K~J Thomas, J~P Hill, M~A Pfeifer, K~Chesnel, Y~Tomioka, Y~Tokura,
  and S~D Kevan.
\newblock Orbital domain dynamics in a doped manganite.
\newblock {\em New Journal of Physics}, 10(5):053023, 2008.

\bibitem{vanhove-1954-pr}
L\'eon Van~Hove.
\newblock Correlations in space and time and born approximation scattering in
  systems of interacting particles.
\newblock {\em Phys. Rev.}, 95:249--262, Jul 1954.

\bibitem{Gutt-2009-OptExp}
C.~Gutt, L.~M. Stadler, A.~Duri, T.~Autenrieth, O.~Leupold, Y.~Chushkin, and
  G.~Gr\"{u}bel.
\newblock Measuring temporal speckle correlations at ultrafast x-ray sources.
\newblock {\em Opt. Express}, 17(1):55--61, 2009.

\bibitem{pellegrini-2016-rmp}
C.~Pellegrini, A.~Marinelli, and S.~Reiche.
\newblock The physics of x-ray free-electron lasers.
\newblock {\em Rev. Mod. Phys.}, 88:015006, Mar 2016.

\bibitem{li-2014-jsr}
Luxi Li, Pawe{\l} Kwa{\'{s}}niewski, Davide Orsi, Lutz Wiegart, Luigi
  Cristofolini, Chiara Caronna, and Andrei Fluerasu.
\newblock {Photon statistics and speckle visibility spectroscopy with partially
  coherent X-rays}.
\newblock {\em Journal of Synchrotron Radiation}, 21(6):1288--1295, Nov 2014.

\bibitem{green-1985-prb}
F.~Green, D.~Neilson, and J.~Szyma\ifmmode~\acute{n}\else \'{n}\fi{}ski.
\newblock First-principles calculation of the dynamic structure factor for the
  electron gas in metallic systems.
\newblock {\em Phys. Rev. B}, 31:5837--5840, May 1985.

\bibitem{goodman-2007-book}
J.W. Goodman.
\newblock {\em Speckle Phenomena in Optics: Theory and Applications}.
\newblock Roberts \& Company, 2007.

\bibitem{goodman-1976-josa}
J.~W. Goodman.
\newblock Some fundamental properties of speckle$\ast$.
\newblock {\em J. Opt. Soc. Am.}, 66(11):1145--1150, Nov 1976.

\bibitem{shen-2020-prb}
L.~Shen, S.~A. Mack, G.~Dakovski, G.~Coslovich, O.~Krupin, M.~Hoffmann, S.-W.
  Huang, Y-D. Chuang, J.~A. Johnson, S.~Lieu, S.~Zohar, C.~Ford, M.~Kozina,
  W.~Schlotter, M.~P. Minitti, J.~Fujioka, R.~Moore, W-S. Lee, Z.~Hussain,
  Y.~Tokura, P.~Littlewood, and J.~J. Turner.
\newblock Decoupling spin-orbital correlations in a layered manganite amidst
  ultrafast hybridized charge-transfer band excitation.
\newblock {\em Phys. Rev. B}, 101:201103, May 2020.

\bibitem{kraus-2017-natastron}
D.~Kraus, J.~Vorberger, A.~Pak, N.~J. Hartley, L.~B. Fletcher, S.~Frydrych,
  E.~Galtier, E.~J. Gamboa, D.~O. Gericke, S.~H. Glenzer, E.~Granados, M.~J.
  MacDonald, A.~J. MacKinnon, E.~E. McBride, I.~Nam, P.~Neumayer, M.~Roth,
  A.~M. Saunders, A.~K. Schuster, P.~Sun, T.~van Driel, T.~D{\"o}ppner, and
  R.~W. Falcone.
\newblock Formation of diamonds in laser-compressed hydrocarbons at planetary
  interior conditions.
\newblock {\em Nature Astronomy}, 1(9):606--611, 2017.

\bibitem{mcbride-2019-natphys}
E.~E. McBride, A.~Krygier, A.~Ehnes, E.~Galtier, M.~Harmand,
  Z.~Kon{\^o}pkov{\'a}, H.~J. Lee, H.~P. Liermann, B.~Nagler, A.~Pelka,
  M.~R{\"o}del, A.~Schropp, R.~F. Smith, C.~Spindloe, D.~Swift, F.~Tavella,
  S.~Toleikis, T.~Tschentscher, J.~S. Wark, and A.~Higginbotham.
\newblock Phase transition lowering in dynamically compressed silicon.
\newblock {\em Nature Physics}, 15(1):89--94, 2019.

\bibitem{seaberg-2017-prl}
M.~H. Seaberg, B.~Holladay, J.~C.~T. Lee, M.~Sikorski, A.~H. Reid, S.~A.
  Montoya, G.~L. Dakovski, J.~D. Koralek, G.~Coslovich, S.~Moeller, W.~F.
  Schlotter, R.~Streubel, S.~D. Kevan, P.~Fischer, E.~E. Fullerton, J.~L.
  Turner, F.-J. Decker, S.~K. Sinha, S.~Roy, and J.~J. Turner.
\newblock Nanosecond x-ray photon correlation spectroscopy on magnetic
  skyrmions.
\newblock {\em Phys. Rev. Lett.}, 119:067403, Aug 2017.

\bibitem{Woo2015}
Seonghoon Woo, Kai Litzius, Benjamin Kr{\"{u}}ger, Mi-young Im, Lucas Caretta,
  Kornel Richter, Maxwell Mann, Andrea Krone, Robert Reeve, Markus Weigand,
  et~al.
\newblock {Observation of room-temperature magnetic skyrmions and their
  current-driven dynamics in ultrathin metallic ferromagnets}.
\newblock {\em Nature Materials}, 15(May):501--507, 2016.

\bibitem{Sergio-PhysRevB}
S.~A. Montoya, S.~Couture, J.~J. Chess, J.~C.~T. Lee, N.~Kent, D.~Henze, S.~K.
  Sinha, M.-Y. Im, S.~D. Kevan, P.~Fischer, et~al.
\newblock Tailoring magnetic energies to form dipole skyrmions and skyrmion
  lattices.
\newblock {\em Phys. Rev. B}, 95:024415, Jan 2017.

\bibitem{Sergio2-PhysRevB}
S.~A. Montoya, S.~Couture, J.~J. Chess, J.~C.~T. Lee, N.~Kent, M.-Y. Im, S.~D.
  Kevan, P.~Fischer, B.~J. McMorran, S.~Roy, V.~Lomakin, and E.~E. Fullerton.
\newblock Resonant properties of dipole skyrmions in amorphous fe/gd
  multilayers.
\newblock {\em Phys. Rev. B}, 95:224405, Jun 2017.

\bibitem{Dakovski-2015-JSR}
Georgi~L. Dakovski, Philip Heimann, Michael Holmes, Oleg Krupin, Michael~P.
  Minitti, Ankush Mitra, Stefan Moeller, Michael Rowen, William~F. Schlotter,
  and Joshua~J. Turner.
\newblock The soft x-ray research instrument at the linac coherent light
  source.
\newblock {\em J. Synchrotron Rad.}, 22(3):498--502, 2015.

\bibitem{Bostedt-2016-RMP}
Christoph Bostedt, S\'ebastien Boutet, David~M. Fritz, Zhirong Huang, Hae~Ja
  Lee, Henrik~T. Lemke, Aymeric Robert, William~F. Schlotter, Joshua~J. Turner,
  and Garth~J. Williams.
\newblock Linac coherent light source: The first five years.
\newblock {\em Rev. Mod. Phys.}, 88:015007, Mar 2016.

\bibitem{Schlotter-2012-RSI}
W.~F. Schlotter, J.~J. Turner, M.~Rowen, P.~Heimann, M.~Holmes, O.~Krupin,
  M.~Messerschmidt, S.~Moeller, J.~Krzywinski, R.~Soufli, et~al.
\newblock The soft x-ray instrument for materials studies at the linac coherent
  light source x-ray free-electron laser.
\newblock {\em Rev. Sci. Instrum.}, 83(4):043107--043107--10, 2012.

\bibitem{Heimann-2011-RSI}
P.~Heimann, O.~Krupin, W.~F. Schlotter, J.~Turner, J.~Krzywinski,
  F.~Sorgenfrei, M.~Messerschmidt, D.~Bernstein, J.~Chalupsky, V.~Hajkova,
  et~al.
\newblock Linac coherent light source soft x-ray materials science instrument
  optical design and monochromator commissioning.
\newblock {\em Rev. Sci. Instrum.}, 82(9):093104--093104--8, 2011.

\bibitem{Tiedtke-2014-OptExpress}
K.~Tiedtke, A.~A. Sorokin, U.~Jastrow, P.~Juranic, S.~Kreis, N.~Gerken,
  M.~Richter, U.~Arp, Y.~Feng, D.~Nordlund, et~al.
\newblock Absolute pulse energy measurements of soft x-rays at the linac
  coherent light source.
\newblock {\em Opt. Express}, 22(18):21214--21226, Sep 2014.

\bibitem{Chalupsky-2011-NIMPR}
J.~Chalupsky, P.~Bohacek, V.~Hajkova, S.~P. Hau-Riege, P.~A. Heimann, L.~Juha,
  J.~Krzywinski, M.~Messerschmidt, S.~P. Moeller, B.~Nagler, et~al.
\newblock Comparing different approaches to characterization of focused x-ray
  laser beams.
\newblock {\em Nucl. Instrum. Methods Phys. Res. A}, 631(1):130--133, 2011.

\bibitem{Krupin-2012-OptExp}
O.~Krupin, M.~Trigo, W.~F. Schlotter, M.~Beye, F.~Sorgenfrei, J.~J. Turner,
  D.~A. Reis, N.~Gerken, S.~Lee, W.~S. Lee, et~al.
\newblock Temporal cross-correlation of x-ray free electron and optical lasers
  using soft x-ray pulse induced transient reflectivity.
\newblock {\em Opt. Express}, 20(10):11396--11406, 2012.

\bibitem{Beye-2012-APL}
M.~Beye, O.~Krupin, G.~Hays, A.~H. Reid, D.~Rupp, S.~de Jong, S.~Lee, W.~S.
  Lee, Y.~D. Chuang, R.~Coffee, et~al.
\newblock X-ray pulse preserving single-shot optical cross-correlation method
  for improved experimental temporal resolution.
\newblock {\em Appl. Phys. Lett.}, 100(12):121108--121108--4, 2012.

\bibitem{zohar-2019-ol}
Sioan Zohar and Joshua~J. Turner.
\newblock Multivariate analysis of x-ray scattering using a stochastic source.
\newblock {\em Opt. Lett.}, 44(2):243--246, Jan 2019.

\bibitem{esposito-2020-apl}
V.~Esposito, X.~Y. Zheng, M.~H. Seaberg, S.~A. Montoya, B.~Holladay, A.~H.
  Reid, R.~Streubel, J.~C.~T. Lee, L.~Shen, J.~D. Koralek, G.~Coslovich,
  P.~Walter, S.~Zohar, V.~Thampy, M.~F. Lin, P.~Hart, K.~Nakahara, P.~Fischer,
  W.~Colocho, A.~Lutman, F.-J. Decker, S.~K. Sinha, E.~E. Fullerton, S.~D.
  Kevan, S.~Roy, M.~Dunne, and J.~J. Turner.
\newblock Skyrmion fluctuations at a first-order phase transition boundary.
\newblock {\em Applied Physics Letters}, 116(18):181901, 2020.

\bibitem{seaberg-2015-apl}
Matthew~H. Seaberg, Alexandre d'Aspremont, and Joshua~J. Turner.
\newblock Coherent diffractive imaging using randomly coded masks.
\newblock {\em Applied Physics Letters}, 107(23):231103, 2015.

\bibitem{waldspurger-2015-mathprog}
Ir{\`e}ne Waldspurger, Alexandre d'Aspremont, and St{\'e}phane Mallat.
\newblock Phase recovery, maxcut and complex semidefinite programming.
\newblock {\em Mathematical Programming}, 149(1):47--81, 2015.

\bibitem{fogel-2016-mpc}
Fajwel Fogel, Ir{\`e}ne Waldspurger, and Alexandre d'Aspremont.
\newblock Phase retrieval for imaging problems.
\newblock {\em Mathematical Programming Computation}, 8(3):311--335, 2016.

\bibitem{Miao-2015-science}
Jianwei Miao, Tetsuya Ishikawa, Ian~K Robinson, and Margaret~M Murnane.
\newblock {Beyond crystallography: Diffractive imaging using coherent x-ray
  light sources}.
\newblock {\em Science}, 348(6234):530--535, 2015.

\bibitem{Lee-2016-APL}
J.~C.~T Lee, J.~J. Chess, S.~A. Montoya, X.~Shi, N.~Tamura, S.~K. Mishra,
  P.~Fischer, B.~J. McMorran, S.~K. Sinha, E.~E. Fullerton, et~al.
\newblock Synthesizing skyrmion bound pairs in fe-gd thin films.
\newblock {\em Appl. Phys. Lett.}, 109(2):022402, 2016.

\bibitem{yu-2011-natmat}
X.~Z. Yu, N.~Kanazawa, Y.~Onose, K.~Kimoto, W.~Z. Zhang, S.~Ishiwata,
  Y.~Matsui, and Y.~Tokura.
\newblock Near room-temperature formation of a skyrmion crystal in thin-films
  of the helimagnet fege.
\newblock {\em Nat Mater}, 10(2):106--109, 02 2011.

\bibitem{phillips-1996-rpp}
J~C Phillips.
\newblock Stretched exponential relaxation in molecular and electronic glasses.
\newblock {\em Reports on Progress in Physics}, 59(9):1133--1207, sep 1996.

\bibitem{cheong-1991-prl}
S-W. Cheong, G.~Aeppli, T.~E. Mason, H.~Mook, S.~M. Hayden, P.~C. Canfield,
  Z.~Fisk, K.~N. Clausen, and J.~L. Martinez.
\newblock Incommensurate magnetic fluctuations in
  ${\mathrm{la}}_{2\mathrm{\ensuremath{-}}\mathit{x}}$${\mathrm{sr}}_{\mathit{x}}$${\mathrm{cuo}}_{4}$.
\newblock {\em Phys. Rev. Lett.}, 67:1791--1794, Sep 1991.

\bibitem{Ghiringhelli-2012-science}
G.~Ghiringhelli, M.~Le~Tacon, M.~Minola, S.~Blanco-Canosa, C.~Mazzoli, N.~B.
  Brookes, G.~M. De~Luca, A.~Frano, D.~G. Hawthorn, F.~He, T.~Loew, M.~Moretti
  Sala, D.~C. Peets, M.~Salluzzo, E.~Schierle, R.~Sutarto, G.~A. Sawatzky,
  E.~Weschke, B.~Keimer, and L.~Braicovich.
\newblock Long-range incommensurate charge fluctuations in
  $\mathrm{(Y,Nd)B}{\mathrm{a}}_{2}\mathrm{C}{\mathrm{u}}_{3}${{O}}$_{6+x}$.
\newblock {\em Science}, 337(6096):821--825, 2012.

\bibitem{achkar-2012-prl}
A.~J. Achkar, R.~Sutarto, X.~Mao, F.~He, A.~Frano, S.~Blanco-Canosa,
  M.~Le~Tacon, G.~Ghiringhelli, L.~Braicovich, M.~Minola, M.~Moretti~Sala,
  C.~Mazzoli, Ruixing Liang, D.~A. Bonn, W.~N. Hardy, B.~Keimer, G.~A.
  Sawatzky, and D.~G. Hawthorn.
\newblock Distinct charge orders in the planes and chains of
  ortho-$\mathrm{III}$-ordered
  $\mathrm{YB}{\mathrm{a}}_{2}\mathrm{C}{\mathrm{u}}_{3}${{O}}$_{6+\delta}$
  superconductors identified by resonant elastic x-ray scattering.
\newblock {\em Phys. Rev. Lett.}, 109:167001, Oct 2012.

\bibitem{chang-2012-natphys}
J.~Chang, E.~Blackburn, A.~T. Holmes, N.~B. Christensen, J.~Larsen, J.~Mesot,
  Ruixing Liang, D.~A. Bonn, W.~N. Hardy, A.~Watenphul, M.~v. Zimmermann, E.~M.
  Forgan, and S.~M. Hayden.
\newblock Direct observation of competition between superconductivity and
  charge density wave order in
  $\mathrm{YB}{\mathrm{a}}_{2}\mathrm{C}{\mathrm{u}}_{3}${{O}}$_{6.67}$.
\newblock {\em Nat Phys}, 8(12):871--876, 2012.

\bibitem{emery-1999-pnas}
V.~J. Emery, S.~A. Kivelson, and J.~M. Tranquada.
\newblock Stripe phases in high-temperature superconductors.
\newblock {\em Proceedings of the National Academy of Sciences},
  96(16):8814--8817, 1999.

\bibitem{dai-2001-prb}
Pengcheng Dai, H.~A. Mook, R.~D. Hunt, and F.~Do\ifmmode~\breve{g}\else
  \u{g}\fi{}an.
\newblock Evolution of the resonance and incommensurate spin fluctuations in
  superconducting
  ${\mathrm{yb}}\mathrm{{a}}_{2}{\mathrm{c}}{\mathrm{u}}_{3}{\mathrm{o}}_{6+x}$.
\newblock {\em Phys. Rev. B}, 63:054525, Jan 2001.

\bibitem{Hufner-2008-rpp}
S~Hüfner, M~A Hossain, A~Damascelli, and G~A Sawatzky.
\newblock Two gaps make a high-temperature superconductor?
\newblock {\em Reports on Progress in Physics}, 71(6):062501, may 2008.

\bibitem{Tranquada2004}
J.~M. Tranquada, H.~Woo, T.~G. Perring, H.~Goka, G.~D. Gu, G.~Xu, M.~Fujita,
  and K.~Yamada.
\newblock Quantum magnetic excitations from stripes in copper oxide
  superconductors.
\newblock {\em Nature}, 429(6991):534--538, Jun 2004.

\bibitem{Vignolle2007}
B.~Vignolle, S.~M. Hayden, D.~F. McMorrow, H.~M. R{\o}nnow, B.~Lake, C.~D.
  Frost, and T.~G. Perring.
\newblock Two energy scales in the spin excitations of the high-temperature
  superconductor LSCO.
\newblock {\em Nature Physics}, 3(3):163--167, Mar 2007.

\bibitem{LeTacon-2011-NP}
M.~Le~Tacon, G.~Ghiringhelli, J.~haloupka, M.~Moretti~Sala, V.~Hinkov, M.~W.
  Haverkort, M.~Minola, M.~Bakr, K.~J Zhou, S.~Blanco-Canosa, C.~Monney, Y.~T.
  Song, G.~L. Sun, C.~T. Lin, G.~M. De~Luca, M.~Salluzzo, G.~Khaliullin,
  T.~Schmitt, L.~Braicovich, and B.~Keimer.
\newblock Intense paramagnon excitations in a large family of high-temperature
  superconductors.
\newblock {\em Nature Physics}, 7:725--730, July 2011.

\bibitem{miao-2017-pnas}
H.~Miao, J.~Lorenzana, G.~Seibold, Y.~Y. Peng, A.~Amorese, F.~Yakhou-Harris,
  K.~Kummer, N.~B. Brookes, R.~M. Konik, V.~Thampy, G.~D. Gu, G.~Ghiringhelli,
  L.~Braicovich, and M.~P.~M. Dean.
\newblock High-temperature charge density wave correlations in
  $\mathrm{L}{\mathrm{a}}_{1.875}\mathrm{B}{\mathrm{a}}_{0.125}{\mathrm{cu}}\mathrm{O}_{4}$
  without spin{\textendash}charge locking.
\newblock {\em Proceedings of the National Academy of Sciences},
  114(47):12430--12435, 2017.

\bibitem{Hepting2018}
M.~Hepting, L.~Chaix, E.~W. Huang, R.~Fumagalli, Y.~Y. Peng, B.~Moritz,
  K.~Kummer, N.~B. Brookes, W.~C. Lee, M.~Hashimoto, T.~Sarkar, J.-F. He, C.~R.
  Rotundu, Y.~S. Lee, R.~L. Greene, L.~Braicovich, G.~Ghiringhelli, Z.~X. Shen,
  T.~P. Devereaux, and W.~S. Lee.
\newblock Three-dimensional collective charge excitations in electron-doped
  copper oxide superconductors.
\newblock {\em Nature}, 563(7731):374--378, Nov 2018.

\bibitem{Arpaia906}
R.~Arpaia, S.~Caprara, R.~Fumagalli, G.~De~Vecchi, Y.~Y. Peng, E.~Andersson,
  D.~Betto, G.~M. De~Luca, N.~B. Brookes, F.~Lombardi, M.~Salluzzo,
  L.~Braicovich, C.~Di~Castro, M.~Grilli, and G.~Ghiringhelli.
\newblock Dynamical charge density fluctuations pervading the phase diagram of
  a cu-based high-tc superconductor.
\newblock {\em Science}, 365(6456):906--910, 2019.

\bibitem{torch-2013-natphys}
Darius~H. Torchinsky, Fahad Mahmood, Anthony~T. Bollinger, Ivan Bo{\v
  z}ovi{\'c}, and Nuh Gedik.
\newblock Fluctuating charge-density waves in a cuprate superconductor.
\newblock {\em Nature Materials}, 12(5):387--391, 2013.

\bibitem{hinton-2013-prb}
J.~P. Hinton, J.~D. Koralek, Y.~M. Lu, A.~Vishwanath, J.~Orenstein, D.~A. Bonn,
  W.~N. Hardy, and Ruixing Liang.
\newblock New collective mode in yba${}_{2}$cu${}_{3}$o${}_{6+x}$ observed by
  time-domain reflectometry.
\newblock {\em Phys. Rev. B}, 88:060508, Aug 2013.

\bibitem{dakovski-2015-prb}
Georgi~L. Dakovski, Wei-Sheng Lee, David~G. Hawthorn, Niklas Garner, Doug Bonn,
  Walter Hardy, Ruixing Liang, Matthias~C. Hoffmann, and Joshua~J. Turner.
\newblock Enhanced coherent oscillations in the superconducting state of
  underdoped
  $\mathrm{YB}{\mathrm{a}}_{2}\mathrm{C}{\mathrm{u}}_{3}{\mathrm{o}}_{6+x}$
  induced via ultrafast terahertz excitation.
\newblock {\em Phys. Rev. B}, 91:220506, Jun 2015.

\bibitem{mitrano-2019-sciadv}
Matteo Mitrano, Sangjun Lee, Ali~A. Husain, Luca Delacretaz, Minhui Zhu,
  Gilberto de~la Pe{\~n}a~Munoz, Stella X.-L. Sun, Young~Il Joe, Alexander~H.
  Reid, Scott~F. Wandel, Giacomo Coslovich, William Schlotter, Tim van Driel,
  John Schneeloch, G.~D. Gu, Sean Hartnoll, Nigel Goldenfeld, and Peter
  Abbamonte.
\newblock Ultrafast time-resolved x-ray scattering reveals diffusive charge
  order dynamics in la2{\textendash}xbaxcuo4.
\newblock {\em Science Advances}, 5(8), 2019.

\bibitem{wadel-2020-arxiv}
S.~Wandel et~al.
\newblock Light-enhanced charge density wave coherence in a high-temperature
  superconductor.
\newblock {\em arXiv:2003.04224 [cond-mat.supr-con]}, 2020.

\bibitem{kivelson-2003-rmp}
S.~A. Kivelson, I.~P. Bindloss, E.~Fradkin, V.~Oganesyan, J.~M. Tranquada,
  A.~Kapitulnik, and C.~Howald.
\newblock How to detect fluctuating stripes in the high-temperature
  superconductors.
\newblock {\em Rev. Mod. Phys.}, 75:1201--1241, Oct 2003.

\bibitem{RevModPhys.87.457}
Eduardo Fradkin, Steven~A. Kivelson, and John~M. Tranquada.
\newblock Colloquium: Theory of intertwined orders in high temperature
  superconductors.
\newblock {\em Rev. Mod. Phys.}, 87:457--482, May 2015.

\bibitem{kivelson-1998-nature}
S.~A. Kivelson, E.~Fradkin, and V.~J. Emery.
\newblock Electronic liquid-crystal phases of a doped mott insulator.
\newblock {\em Nature}, 393:550 EP --, 06 1998.

\bibitem{chen-2016-prl}
X.~M. Chen, V.~Thampy, C.~Mazzoli, A.~M. Barbour, H.~Miao, G.~D. Gu, Y.~Cao,
  J.~M. Tranquada, M.~P.~M. Dean, and S.~B. Wilkins.
\newblock Remarkable stability of charge density wave order in
  ${\mathrm{la}}_{1.875}{\mathrm{ba}}_{0.125}{\mathrm{cuo}}_{4}$.
\newblock {\em Phys. Rev. Lett.}, 117:167001, Oct 2016.

\bibitem{PhysRevB.101.174508}
P.~M. Singer, A.~Arsenault, T.~Imai, and M.~Fujita.
\newblock $^{139}\mathrm{La}$ nmr investigation of the interplay between
  lattice, charge, and spin dynamics in the charge-ordered high-${T}_{c}$
  cuprate ${\mathrm{la}}_{1.875}{\mathrm{ba}}_{0.125}{\mathrm{cuo}}_{4}$.
\newblock {\em Phys. Rev. B}, 101:174508, May 2020.

\bibitem{markiewicz-2016-scirep}
R.~S. Markiewicz, J.~Lorenzana, G.~Seibold, and A.~Bansil.
\newblock Short range smectic order driving long range nematic order: example
  of cuprates.
\newblock {\em Scientific Reports}, 6:19678 EP --, 01 2016.

\bibitem{price-1999-prl}
A.~C. Price, L.~B. Sorensen, S.~D. Kevan, J.~Toner, A.~Poniewierski, and
  R.~Ho\l{}yst.
\newblock Coherent soft-x-ray dynamic light scattering from smectic-
  $\mathit{A}$ films.
\newblock {\em Phys. Rev. Lett.}, 82:755--758, Jan 1999.

\bibitem{markiewicz-scirep-2017}
R.~S. Markiewicz, I.~G. Buda, P.~Mistark, C.~Lane, and A.~Bansil.
\newblock Entropic origin of pseudogap physics and a mott-slater transition in
  cuprates.
\newblock {\em Scientific Reports}, 7:44008 EP --, 03 2017.

\bibitem{Soufli-2012-AppOpt}
Regina Soufli, Mónica Fernández-Perea, Sherry~L Baker, Jeff~C Robinson,
  Eric~M Gullikson, Philip Heimann, Valeriy~V Yashchuk, Wayne~R McKinney,
  William~F Schlotter, and Michael Rowen.
\newblock Development and calibration of mirrors and gratings for the soft
  x-ray materials science beamline at the linac coherent light source
  free-electron laser.
\newblock {\em Appl. Opt.}, 51(12):2118--2128, 2012.

\bibitem{Moeller-2015-JSR}
Stefan Moeller, Garth Brown, Georgi Dakovski, Bruce Hill, Michael Holmes,
  Jennifer Loos, Ricardo Maida, Ernesto Paiser, William Schlotter, Joshua~J.
  Turner, et~al.
\newblock Pulse energy measurement at the $\mathrm{SXR}$ instrument.
\newblock {\em J. Synchrotron Rad.}, 22(3):606--611, 2015.

\bibitem{dainty-1985-book}
J.C. Dainty.
\newblock {\em Laser Speckle and Related Phenomena}.
\newblock Springer-Verlag Berlin Heidelberg, 1985.

\bibitem{Hruszkewycz-2012-PRL}
S.~O. Hruszkewycz, M.~Sutton, P.~H. Fuoss, B.~Adams, S.~Rosenkranz, Jr. Ludwig,
  K.~F., W.~Roseker, D.~Fritz, M.~Cammarata, D.~Zhu, et~al.
\newblock High contrast x-ray speckle from atomic-scale order in liquids and
  glasses.
\newblock {\em Phys. Rev. Lett.}, 109(18):185502, 2012.

\bibitem{Sikorski2016}
Marcin Sikorski, Yiping Feng, Sanghoon Song, Diling Zhu, Gabriella Carini, Sven
  Herrmann, Kurtis Nishimura, Philip Hart, and Aymeric Robert.
\newblock {Application of an ePix100 detector for coherent scattering using a
  hard X-ray free-electron laser}.
\newblock {\em J. Synchrotron Rad.}, 23:1171--1179, 2016.

\end{thebibliography}
\end{document}